\documentclass[aps,prx,reprint, superscriptaddress]{revtex4-1}
\usepackage[english]{babel}
\usepackage{amsmath}
\usepackage{amsfonts}
\usepackage{amssymb}
\usepackage{graphicx}
\usepackage{bm}
\usepackage{bbm}
\usepackage{braket}
\usepackage{color, ulem, mathtools}
\AtBeginDocument{%
	\newwrite\bibnotes
	\def\bibnotesext{Notes.bib}
	\immediate\openout\bibnotes=\jobname\bibnotesext
	\immediate\write\bibnotes{@CONTROL{REVTEX41Control}}
	\immediate\write\bibnotes{@CONTROL{%
			apsrev41Control,author="08",editor="1",pages="1",title="0",year="1"}}
	\if@filesw
	\immediate\write\@auxout{\string\citation{apsrev41Control}}%
	\fi
}%

\begin{document}
\title{Coherent transfer of quantum information in silicon using resonant SWAP gates}

\author{A. J. Sigillito}
\affiliation{Department of Physics, Princeton University, Princeton, New Jersey 08544, USA}
\author{M. J. Gullans}
\affiliation{Department of Physics, Princeton University, Princeton, New Jersey 08544, USA}
\author{L. F. Edge}
\affiliation{HRL Laboratories LLC, 3011 Malibu Canyon Road, Malibu, California 90265, USA}
\author{M. Borselli}
\affiliation{HRL Laboratories LLC, 3011 Malibu Canyon Road, Malibu, California 90265, USA}
\author{J. R. Petta}
\affiliation{Department of Physics, Princeton University, Princeton, New Jersey 08544, USA}


\renewcommand{\topfraction}{0.9}	
\renewcommand{\bottomfraction}{0.8}	
\setcounter{topnumber}{2}
\setcounter{bottomnumber}{2}
\setcounter{totalnumber}{4}     
\setcounter{dbltopnumber}{2}    
\renewcommand{\dbltopfraction}{0.7}	
\renewcommand{\textfraction}{0.027}	
\renewcommand{\floatpagefraction}{0.8}	
\renewcommand{\dblfloatpagefraction}{0.7}	
\date{\today}

\maketitle

\textbf{
Solid state quantum processors based on spins in silicon quantum dots are emerging as a powerful platform for quantum information processing \cite{loss1998,Zajac2018,Watson2018}. High fidelity single- and two-qubit gates have recently been demonstrated \cite{Veldhorst2015,Zajac2018,Watson2018,Huang2018,Xue2018} and large extendable qubit arrays are now routinely fabricated \cite{Mills2018,SigillitoQuadDot}. However, two-qubit gates are mediated through nearest-neighbor exchange interactions \cite{loss1998,petta2005}, which require direct wavefunction overlap. This limits the overall connectivity of these devices and is a major hurdle to realizing error correction \cite{Fowler2004}, quantum random access memory \cite{qram}, and multi-qubit quantum algorithms \cite{QAOA}. To extend the connectivity, qubits can be shuttled around a device using quantum SWAP gates, but phase coherent SWAPs have not yet been realized in silicon devices \cite{Veldhorst2015, Zajac2018, Watson2018, Huang2018, Xue2018}. Here, we demonstrate a new single-step resonant SWAP gate.  We first use the gate to efficiently initialize and readout our double quantum dot. We then show that the gate can move spin eigenstates in 100 ns with average fidelity ${\bar{\bm{F}}_{\rm \textbf{SWAP}}^{\bm{(p)}}}=\bm{98\, \%}$. Finally, the transfer of arbitrary two-qubit product states is benchmarked using state tomography and Clifford randomized benchmarking \cite{Magesan2012,Huang2018}, yielding an average fidelity of ${\bar{\bm{F}}_{\rm \textbf{SWAP}}^{\bm{(c)}}}=\bm{84\, \%}$ for gate operation times of $\sim${300~ns}. Through coherent spin transport, our resonant SWAP gate enables the coupling of non-adjacent qubits, thus paving the way to large scale experiments using silicon spin qubits.}

\begin{figure}
	\begin{center}
		\includegraphics[width=\columnwidth]{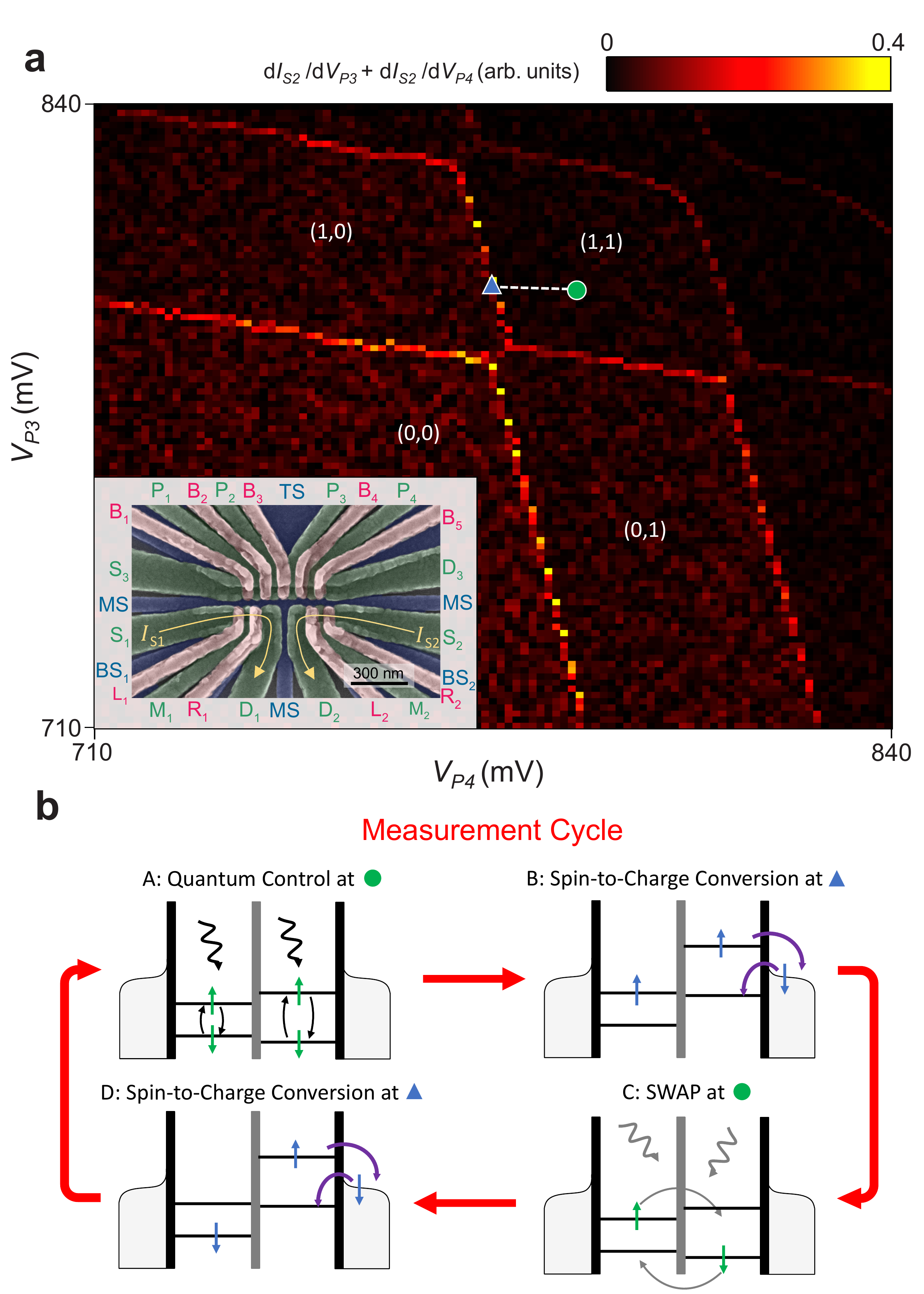}
		\caption{
			(a) Charge stability map for a DQD formed using sites 3 and 4 in the quadruple dot array (inset). Quantum control is performed near the center of the (1,1) charge stability region as denoted by the green circle. Readout of dot 4 is performed at the (1,0)-(1,1) charge transition denoted by the blue triangle. (b) The typical measurement cycle is shown for controlling and reading out two quantum dots. In panel A, the qubits are manipulated and in panel B $Q_4$ is read out through spin-selective tunneling --- leaving the qubit in the $\ket{\downarrow}$ state. In panel C, the exchange interaction $J_{34}$ between $Q_3$ and $Q_4$ is modulated (through modulation of the barrier) to induce a SWAP operation, thus mapping the state of $Q_3$ onto $Q_4$. $Q_4$ is then read out once more to determine the state of $Q_3$.}
		\label{fig1}
	\end{center}	
	\vspace{-0.6cm}
\end{figure}

In this work, we use two sites of a quadruple quantum dot fabricated on a $^{28}$Si/SiGe heterostructure [inset of Fig.~\ref{fig1}(a)] \cite{SigillitoQuadDot}. Electric dipole spin resonance (EDSR) \cite{pioro2008electrically,Tokura2006} enables single-spin control and an on-chip micromagnet detunes the frequency of each spin to enable site-selective control \cite{Yoneda2015, SigillitoQuadDot}. For demonstration purposes, we use two dots in the device with qubits accumulated under plunger gates $P_{3}$ and $P_{4}$. We hereafter refer to the two qubits as $Q_3$ and $Q_4$, respectively. The charge stability diagram of this DQD is shown in Fig.~\ref{fig1}(a) and quantum control is performed in the $(N_{i},N_{i+1}) = (1,1)$ charge configuration, where $N_i$ denotes the number of electrons on dot $i$. We measure the state of $Q_4$ through spin-selective tunneling to a drain reservoir accumulated beneath gate $D_3$ \cite{elzerman2004single}.

There are two modes of operation for the resonant SWAP gate demonstrated in this Letter. First, a \textit{projection}-SWAP can be used to transfer spin eigenstates between quantum dots. The \textit{projection}-SWAP enables rapid initialization and readout of inner sites in an array that are not directly connected to the leads --- meaning they can not be directly initialized or measured. Secondly, a \textit{coherent}-SWAP can be used to transfer arbitrary quantum states between quantum dots, thus, allowing the rearrangement of qubits in the array. A \textit{coherent}-SWAP is crucial for performing multi-qubit algorithms or error correction in devices with limited qubit-to-qubit connectivity \cite{Linke17}. The \textit{coherent}- and \textit{projection}-SWAP gates are realized using the same resonant SWAP gate, however, the \textit{coherent}-SWAP requires more stringent calibration.

In our device architecture, two-qubit gates are mediated through the exchange interaction $J_{i,i+1}(V_{Bi+1})$,  which is proportional to the wavefunction overlap between the two adjacent qubits $i$ and $i+1$.  This wavefunction overlap, and thus the exchange interaction, is controlled by adjusting the barrier gate voltage $V_{Bi+1}$. Here, we realize our SWAP gate through a resonant drive on that barrier at a frequency $f_{\text{SWAP}}$ according to the formula $V_{Bi+1}(t) = V_{Bi+1}^{(dc)} + V_{Bi+1}^{(ac)}\cos (2\pi f_{\text{SWAP}}t+\phi)$ where $V_{Bi+1}^{(ac)}$ is the amplitude of the drive and $V_{Bi+1}^{(dc)}$ is a static offset voltage.

In the absence of a magnetic field gradient, when $J_{i,i+1}\gg \gamma_e |B_i^{\text{tot}} - B_{i+1}^{\text{tot}}|$, where $\gamma_e$ is the electronic gyromagnetic ratio and $B_{i}^{\text{tot}}$ is the magnetic field at dot $i$, the two qubits directly undergo SWAP oscillations \cite{petta2005, Nowack2011,maune2012,Kandel2019}. However, state-of-the-art devices rely on large magnetic field gradients \cite{Zajac2018,Watson2018,SigillitoQuadDot,Xue2018} and our device has  $\gamma_eB_3^{\text{tot}}$ = 16.949 GHz and $\gamma_eB_4^{\text{tot}}$ = 17.089 GHz at an external magnetic field of 410 mT. In this regime, the exchange interaction leads to a CPHASE-like evolution \cite{Watson2018,Meunier2011,Russ2018}. To recover the two-qubit SWAP oscillations in the presence of such large magnetic field gradients, we effectively rotate out the gradient by applying an exchange pulse that is resonant with the difference frequency of the two qubits ($2\pi f_{\text{SWAP}}=\gamma_e |B_i^{\text{tot}}-B_{i+1}^{\text{tot}}|$) \cite{Nichol2017}. This can be qualitatively understood as stroboscopically applying exchange whenever the evolution due to the magnetic field gradient returns to its initial state. 

Resonant modulation of $V_{Bi+1}$ at the difference frequency of the two qubits will drive Rabi rotations in the $\lvert \phi_3, \phi_4 \rangle \in \{ \lvert \uparrow \downarrow \rangle, \lvert \downarrow \uparrow \rangle \}$ subspace, while leaving the fully spin-polarized states unaffected. A $\pi$-pulse in this subspace is a SWAP gate up to additional phases on the state of each qubit \cite{SOM}. These phases do not affect the operation of the \textit{projection}-SWAP, but will affect the \textit{coherent}-SWAP. The procedure for calibrating out the phases is outlined in the supplementary material \cite{SOM}. Our resonant SWAP gate is therefore efficiently realized through a single RF burst on the barrier gate between qubits $i$ and $i+1$.

We first describe how the \textit{projection}-SWAP gate can be used for readout of an interior site in our device using the protocol outlined in Fig.~\ref{fig1}(b). In a typical measurement cycle, after the quantum control is performed at (1,1), $Q_4$ is read out through spin-selective tunneling to the leads [blue triangle in Fig.~\ref{fig1}(a)]. If $Q_4$ is in the $\ket{\uparrow}$ state, the electron has enough energy to tunnel off of the dot and is replaced by a lower energy electron in the $\ket{\downarrow}$ state. However, if $Q_4$ is in the $\ket{\downarrow}$ state, no tunneling occurs \cite{elzerman2004single}. Any charge hops are detected by monitoring the current through an adjacent charge sensor [$I_{S2}$ in Fig.~\ref{fig1}(a)]. If a charge hopping event is detected, we record the spin state as $\ket{\uparrow}$. Regardless of its initial state, $Q_4$ is left in the $\ket{\downarrow}$ state after measurement. We next tune the device back into the (1,1) charge configuration and apply the resonant SWAP gate. In this process, $Q_3$'s state is mapped on to $Q_{4}$ and $Q_3$ is left in the $\ket{\downarrow}$ state. We once again measure $Q_4$ to infer $Q_3$'s original state. This measurement protocol leaves the DQD in the $\ket{\downarrow\downarrow}$ state.

\begin{figure}
	\includegraphics[width=\columnwidth]{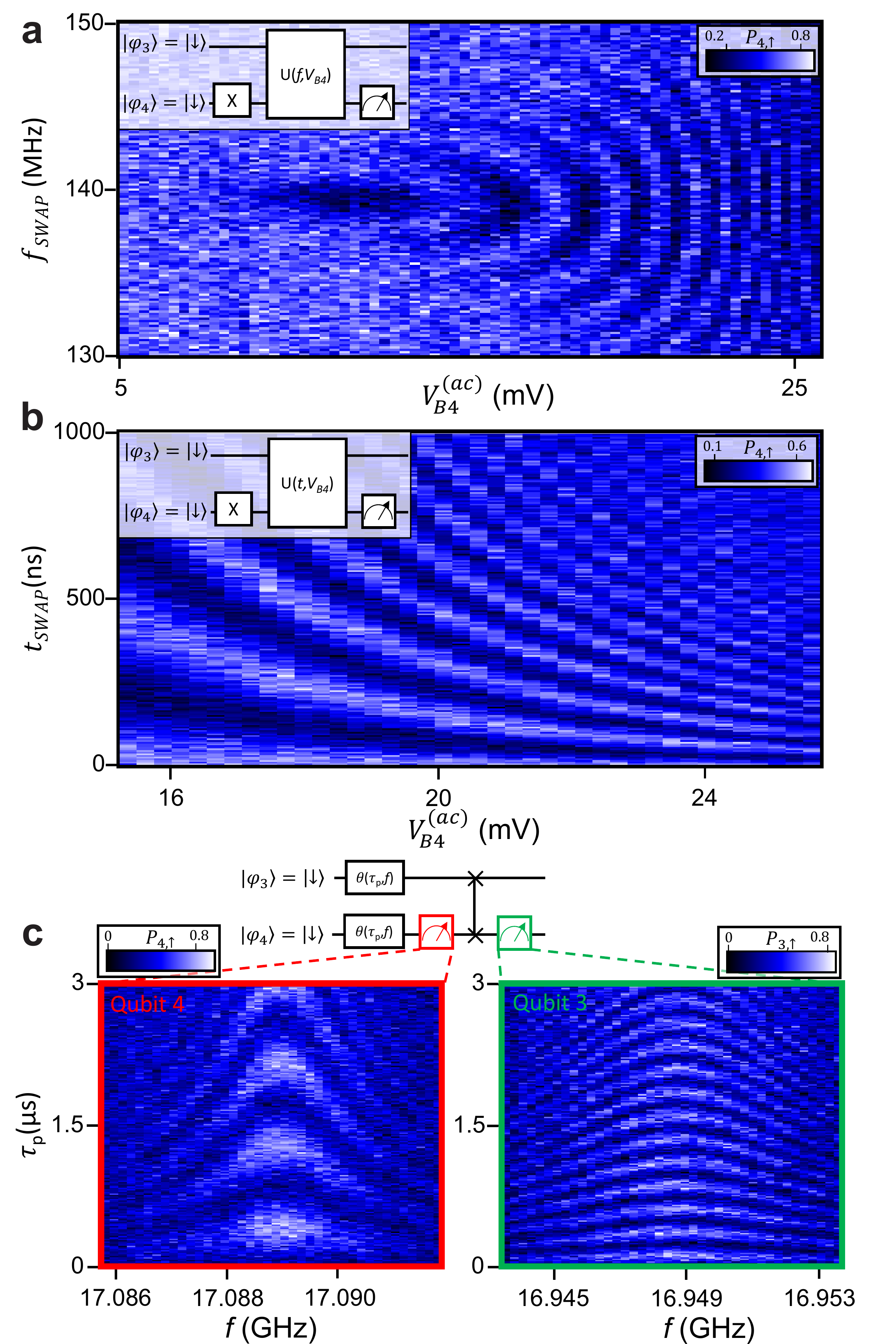}
	\caption{	
		(a) SWAP oscillations measured on $Q_4$, where $P_{4,\uparrow}$ is the probability of measuring $Q_4$ in the spin-up state. The pulse length ($t_{\text{SWAP}}$) is fixed at 600 ns, whereas the amplitude of the drive ($V_{B4}^{(ac)}$) and the frequency of the modulation ($f_{\text{SWAP}}$) are varied to map out the ideal SWAP gate parameters. (b) SWAP oscillations are shown for varying $V_{B4}^{(ac)}$ keeping $f_{\text{SWAP}}$ = 140~MHz. (c) SWAP-based readout is demonstrated for $Q_3$ and $Q_4$, where we simultaneously drive Rabi oscillations on both qubits and then read them out in series. The higher frequency Rabi oscillations observed for $Q_3$ are attributed to a larger transverse field gradient at $Q_3$ \cite{SigillitoQuadDot}.}
	\label{fig2}
	\vspace{-0.6cm}
\end{figure}

To calibrate the SWAP gate, the system is initialized in the $\ket{\phi_3,\phi_4} = \ket{\downarrow\downarrow}$ state through spin-dependent tunneling from a Fermi reservoir into $Q_4$ \cite{Zajac2018}. The SWAP is implemented by driving gate $B_4$ with an RF burst at frequency $f_{\text{SWAP}}$ and duration 600~ns. Figure~\ref{fig2}(a) shows the spin-up probability of $Q_4$, $P_{4,\uparrow}$ as a function of $f_{\text{SWAP}}$ and $V_{B4}^{(ac)}$. For small $V_{B4}^{(ac)}$, there are no measureable SWAP oscillations. At around $V_{B4}^{(ac)}$ = 10~mV, coherent SWAP oscillations in $P_{4,\uparrow}$ appear. The pattern is symmetric about $f_{\text{SWAP}}$ = 140~MHz. For a 600~ns burst at 140~MHz, a SWAP is achieved with $V_{B4}^{ac}$ = 10~mV. To minimize the SWAP time, we fix $f_{\text{SWAP}}$ = 140~MHz and vary $V_{B4}^{(ac)}$ and the drive time ($t_{\text{SWAP}}$) in Fig.~\ref{fig2}(b). Each alternating bright fringe corresponds to an even number of SWAPs. The minimum SWAP time shown here is 23~ns and is limited by the dynamic range of our control electronics. The coherence times $T_2^*$ are approximately 10~$\mu$s for both dots \cite{SigillitoQuadDot}, which is long relative to these gate operation times.

We now demonstrate simultaneous quantum control, initialization, and readout of both dots using spin-to-charge conversion of only $Q_4$. Starting in the $\ket{\downarrow\downarrow}$ state, we apply a microwave burst with duration $\tau_p$ and frequency $f$. We measure $Q_4$ through spin-selective tunneling to $D_3$, leaving $Q_4$ in the $\ket{\downarrow}$ state. We then apply a \textit{projection}-SWAP to the qubits, mapping the state of $Q_3$ onto $Q_4$, and leaving $Q_3$ in the $\ket{\downarrow}$ state. $Q_4$ is then measured so that we can infer the state of $Q_3$. Once $Q_4$ is measured, the qubits are left in the $\ket{\downarrow}$ state, and the DQD is prepared for the next experiment. The spin-up probability for both qubits is plotted as a function of $\tau_p$ and $f$ in Fig.~\ref{fig2}(c). The fringes observed are Rabi oscillations, whose spacing is largest when the qubits are on resonance. These data reveal a qubit difference frequency of 140~MHz, which is consistent with the two qubit spectroscopy in Fig.~\ref{fig2}(a). These data show that we can initialize, control, and readout our DQD even though readout only occurs on $Q_4$.

\begin{figure}[t]
	\begin{center}
		\includegraphics[width=\columnwidth]{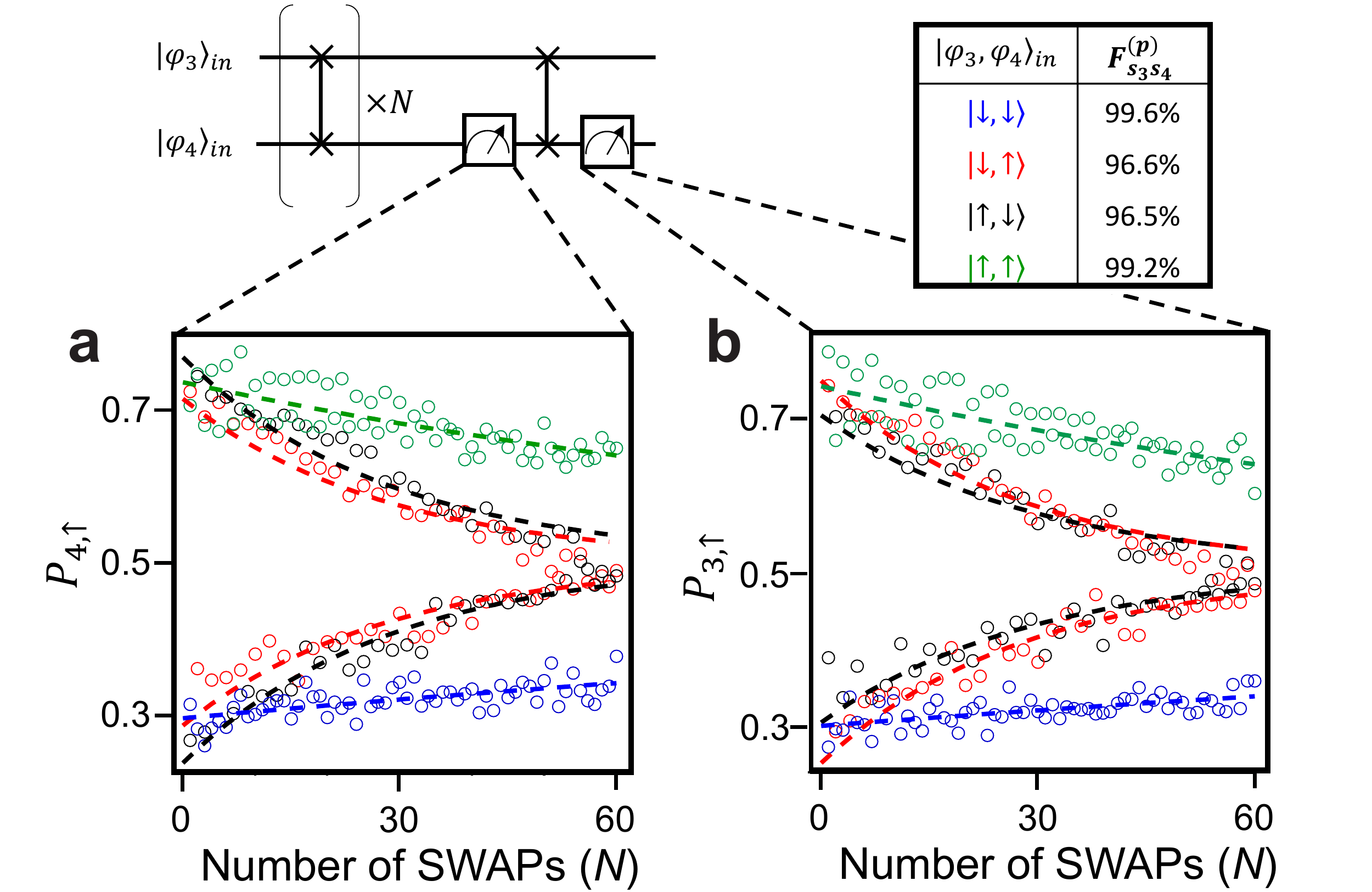}
		\caption{
			The probability of measuring a spin-up $P_{i,\uparrow}$ on $Q_i$ is plotted for $Q_4$ (a) and $Q_3$ (b) after the two qubit system undergoes $N$ SWAP operations. The input states are denoted by color and the dashed lines are fits to the data \cite{SOM}. The extracted fidelity $F_{s_3 s_4}^{(p)}$ from each fit is shown in the table.
		}
		\label{Fig3}
	\end{center}
	\vspace{-0.6cm}
\end{figure}

To quantitatively study the \textit{projection}-SWAP gate for readout purposes, we designed an experiment to be insensitive to state preparation and measurement (SPAM) errors. We prepare the qubits in one of the four spin eigenstates  $\ket{\phi_{3},\phi_{4}}_{in} = \ket{\downarrow\downarrow}$, $\ket{\downarrow\uparrow}$, $\ket{\uparrow\downarrow}$, and $\ket{\uparrow\uparrow}$ before applying the SWAP gate to the qubits $N$ times as shown in Fig.~\ref{Fig3}. We expect the spin polarized states to decay towards a mixed state with $P_{3,\uparrow}$ and $P_{4,\uparrow}$ = 0.5 for large $N$. The antiparallel spin input states should flip-flop between the $\ket{\downarrow\uparrow}$ and $\ket{\uparrow\downarrow}$ states for each additional SWAP we apply. $P_{3,\uparrow}$ and $P_{4,\uparrow}$ will then converge to 0.5 in the large $N$ limit. The decay envelope is given by $F^{(p) N}_{s_3 s_4}$, where $F^{(p)}_{s_3 s_4}$ is the fidelity of the \textit{projection}-SWAP on spin states $s_3$ and $s_4$ between $Q_3$ and $Q_4$. Fitting these curves we find an average fidelity of $F^{(p)}_{ \downarrow \uparrow} = F^{(p)}_{\uparrow \downarrow} = 96.5\%$ for $\ket{\phi_{3},\phi_{4}}_{in} = \ket{\downarrow\uparrow}$ or $\ket{\uparrow\downarrow}$. In cases where both qubits have the same initial state $\ket{\phi_{3},\phi_{4}}_{in} = \ket{\downarrow\downarrow}$ or $\ket{\uparrow\uparrow}$, we achieve fidelities of $F^{(p)}_{\downarrow \downarrow}=99.6\%$ and $F^{(p)}_{\uparrow \uparrow}=99.2\%$ respectively.  Thus, we find an average fidelity for the \textit{projection}-SWAP of $\bar{F}_{\rm SWAP}^{(p)} = 98\, \%$. The spin-polarized input states are insensitive to errors due to noise in the drive field and pulse miscalibrations, but should be sensitive to spin relaxation. The 96.5\% fidelity for antiparallel spin input states is, therefore, not likely limited by relaxation. This is expected, since $T_1$ is 134~ms (52~ms) for $Q_3$ ($Q_4$), which gives an upper bound of 99.97\% fidelity for implementing 60 SWAPs, each padded with a 100 ns idle.  The likely source of errors for antiparallel spin states arises from time-dependent fluctuations in the magnetic field gradients or exchange interaction, which lead to miscalibrations in the \textit{projection}-SWAP gate.

The high-fidelity of our \textit{projection}-SWAP gates implies that this state preparation and measurement technique can be extended to much larger arrays than the DQD configuration studied here. Our current fidelity suggests we could shuttle a spin projection across a nine-dot array \cite{Mills2018} with a fidelity of 85\%. It is notable that while the SWAP gate leads to spin transport between adjacent dots, the electron wavefunctions remain localized on the dots and there is no charge transport. This is in contrast to typical spin-shuttle experiments \cite{baart2016single} that physically move electrons, a process which can be complicated by low-lying valley states \cite{zwanenburg2013silicon}, spin-orbit coupling \cite{Tyryshkin2005}, or spin-relaxation hotspots \cite{Watson2018}. Because spin transport can be controlled using only barrier gates, no fast plunger gate control is necessary, which should enable the operation of devices having fixed charge configurations. This could be of particular interest in two-dimensional arrays where for even small numbers of qubits, charge state control and readout of interior sites becomes unmanageable. Finally, the \textit{projection}-SWAP is compatible with singlet-triplet readout \cite{PHCenhancedlatching,petta2005,Kandel2019} and cavity-based dispersive readout \cite{Zheng2019,Petersson2012}.

\begin{figure}
	\begin{center}
		\includegraphics[width=\columnwidth]{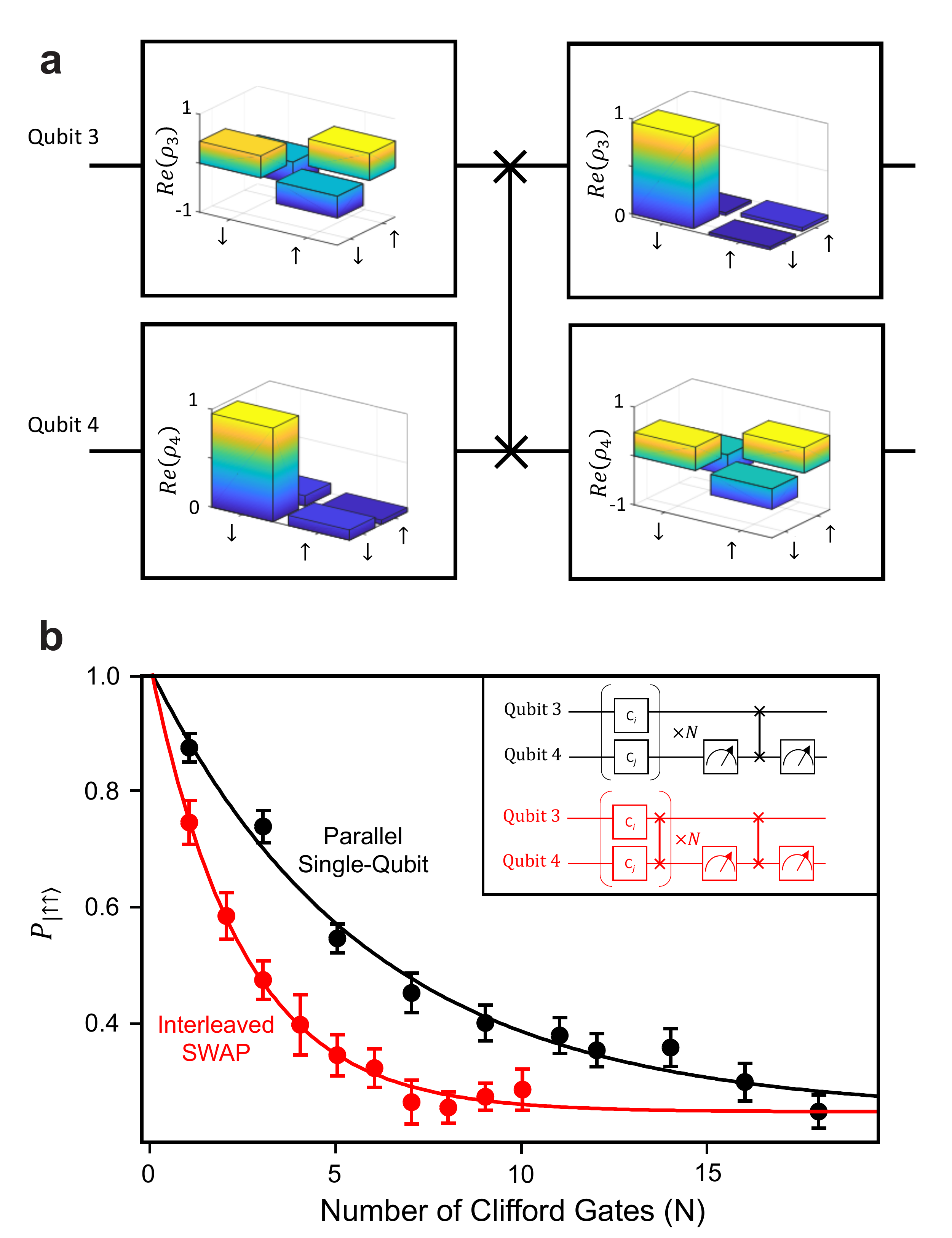}
		\caption{(a) Real part of the density matrix $\rho_i$ for $Q_i$ obtained via state tomography on $Q_3$ and $Q_4$ before (left) and after (right) applying a 302 ns SWAP gate. $Q_3$ was prepared in a superposition state by applying a $\pi$/2 pulse and $Q_4$ is prepared in the down state. After applying the SWAP gate, we find that the density matrices are transferred between the two qubits. The density matrices are corrected for SPAM errors \cite{Watson2018}. $Q_3$ was measured using the \textit{projection}-SWAP protocol outlined in this Letter. From these data we estimate a SWAP fidelity of 89\%. (b) Single qubit randomized benchmarking was performed simultaneously on $Q_3$ and $Q_4$ and the probability of measuring $\ket{\uparrow\uparrow}$ (corrected for SPAM errors) is shown in black as a function of the number of Clifford gates ($N$). The experiment is repeated interleaving SWAP gates with each pair of Clifford gates and the decay is shown in red. Each point consists of 40 random sequences and 500 averages. Error bars denote the standard error. The extracted SWAP fidelity is  $\bar{F}_{\rm SWAP}^{(c)}=84\%$.}
		\label{Fig4}
	\end{center}
	\vspace{-0.6cm}
\end{figure}

With the \textit{projection}-SWAP, we have shown that it is possible to transfer spin eigenstates oriented along the magnetic field axis.  More generally, the ability to transfer arbitrary quantum information is crucial to operating multi-qubit devices with limited connectivity. Therefore, having achieved a high-fidelity \textit{projection}-SWAP, we turn our attention to transferring  product states oriented along arbitrary directions with the \textit{coherent}-SWAP, e.g.,$
    (\alpha \ket{\uparrow} + \beta \ket{\downarrow}) \otimes \ket{\downarrow} \to \ket{\downarrow} \otimes (\alpha \ket{\uparrow} + \beta \ket{\downarrow})$. The \textit{coherent}-SWAP has additional calibration requirements outlined in the methods section. Here we realize a \textit{coherent}-SWAP in 302 ns, which can be made faster by superimposing a dc exchange pulse as outlined in the supplement \cite{SOM}

To verify our calibration, we prepared $Q_3$ in a superposition state, and performed state tomography before and after applying a \textit{coherent}-SWAP. By measuring the $x$, $y$, and $z$ spin projections, we are able to reconstruct the single-spin density matrices $\rho_{i}$ as plotted in Fig.~\ref{Fig4}(a). The imaginary components of $\rho_i$ are shown in the supplementary information \cite{SOM}. From these data, we can estimate the SWAP fidelity $F(\rho)$ by comparing the output state to the targeted state $\psi_{ideal}$ using $F(\rho) = \bra{\psi_{ideal}}\rho\ket{\psi_{ideal}}$ and $\rho =\rho_{3} \otimes \rho_{4}$ \cite{Uhlmannfidelity}. When constructing the two-qubit density matrices, SPAM errors are subtracted out \cite{Watson2018,Zajac2018,SOM}. This analysis gives a state fidelity of $F(\rho)=89\%$. Because this technique only measures the fidelity of swapping one pair of input states, obtaining an average gate fidelity requires repeating the experiment for each possible input.

To measure the average SWAP fidelity, we turn to Clifford randomized benchmarking, which is insensitive to SPAM errors \cite{Knill2008,Huang2018,Magesan2012,Xue2018}. In Clifford randomized benchmarking, quantum circuits consisting of $N$ randomly chosen Clifford gates are applied to a qubit, and at the end of the sequence, the qubit is rotated into a known state. Any gate infidelity throughout the sequence leads to errors in the final state. The qubit is measured and the experiment is repeated varying $N$. As $N$ increases, integrated errors cause the qubit state to become mixed and the probability of measuring $P_{i,\uparrow}$ approaches 50\%. 

To avoid the extensive overhead associated with full two-qubit randomized benchmarking, which requires calibration of an entangling two-qubit gate in addition to the SWAP gate, we use a technique pioneered by Chen \textit{et al.}~\cite{Chen2014} to benchmark two-qubit gates using interleaved randomized benchmarking \cite{Magesan2012,Xue2018}. We first perform single qubit Clifford randomized benchmarking on $Q_3$ and $Q_4$ by measuring the spin-up probability $P_{\ket{\uparrow\uparrow}}$ as a function of sequence length $N$. These data, shown in black in Fig.\ \ref{Fig4}(b), are acquired with $Q_3$ and $Q_4$ single qubit rotations implemented in parallel. We next repeat the randomized benchmarking experiment, this time interleaving $coherent$-SWAP gates after each set of parallel single qubit Clifford gates on $Q_3$ and $Q_4$. These results are plotted in red in Fig.\ \ref{Fig4}(b). The decays are fit to $P_{\ket{\uparrow\uparrow}}=A_{0}p_{c}^m + C$ where $A_{0}$ is the measurement visibility, $C$ is the dark count, and $p_{c}$ is a decay parameter \cite{Magesan2012}. This fit yields a decay parameter $p_c$ = 0.843 for the reference curve and $\bar{p}_c$ = 0.665 for the interleaved curve. By comparing these decay parameters we can extract an average \textit{coherent}-SWAP fidelity of $\bar{F}_{\rm SWAP}^{(c)} = 1-3/4(1-\bar{p_c}/p_c) = 84\%$, which is in good agreement with our estimate from state tomography \cite{Magesan2012}.

In conclusion, we have demonstrated a resonant SWAP gate that can be used for coherent spin transport and high-fidelity state preparation and readout in an array of quantum dot spin qubits. We measure an average \textit{projection}-SWAP gate fidelity of $\bar{F}_{\rm SWAP}^{(p)}=98\%$ when transferring eigenstates with a 100~ns gate time. We further show that a \textit{coherent}-SWAP gate can be used to transfer arbitrary two-qubit states between spins with an average fidelity of $\bar{F}_{\rm SWAP}^{(c)}=84\%$ in $\sim$300 ns as measured using interleaved randomized benchmarking. By implementing automatic calibration and feedback \cite{Huang2018}, we should be able to significantly improve this fidelity. SWAP gates are a crucial building block in any quantum processor with limited qubit-to-qubit connectivity and are necessary to unlock the full capabilities of the multi-qubit devices currently being fabricated in Si/SiGe \cite{SigillitoQuadDot,Mills2018}. This robust implementation of a resonant SWAP gate promises to enable beyond nearest-neighbor operation in quantum dot arrays, which is necessary for quantum information processing with more than two qubits.
\section{methods}
Beyond the calibration required for the \textit{projection}-SWAP, there are three additional constraints that must be satisfied to achieve high fidelity \textit{coherent}-SWAP gates. First, the resonant SWAP pulse must remain phase coherent with the qubits in their doubly-rotating reference frame between calibrations (i.e.\ for hours). Second, because of the constraint that the exchange interaction is always positive, the time-averaged exchange pulse necessarily has some static component, which leads to evolution under an Ising interaction \cite{SOM}. These Ising phases must be calibrated out. Finally, voltage pulses on any gate generally displace both electrons by some small amount. This movement induces phase shifts in both qubits, since they are located in a large magnetic field gradient. These phase shifts must be compensated for.

To satisfy these additional tuning requirements, we first ensure that our RF exchange pulse remains phase coherent. Each qubit's reference frame is defined by the microwave signal generator controlling it, so by mixing together the local oscillators of these signal generators, we obtain a beat frequency that is phase locked to the doubly-rotating two-qubit reference frame. We then amplitude modulate this signal to generate our exchange pulses. A detailed schematic is shown in the supplementary information \cite{SOM}. 

To calibrate for the single and two-qubit Ising phases, we use state tomography on both qubits before and after applying a SWAP gate. In these measurements, we vary the input states to distinguish between errors caused by two-qubit Ising phases, and the single qubit phase shifts. We choose a SWAP time and amplitude such that the Ising phases cancel out, which for this particular configuration occurs for a 302 ns SWAP gate. The single qubit phase shifts were measured to be 180$^{\circ}$ for $Q_3$ and 140$^{\circ}$ for $Q_4$. By superimposing the SWAP pulse with a dc exchange pulse, one can compensate for the Ising phases at arbitrary SWAP lengths, leading to faster operation \cite{SOM}.

\section{Author Contributions}

AJS, MJG, and JRP designed and planned the experiments, AJS fabricated the device and performed the measurements, MJG provided theory support, LFE and MB provided the isotopically enriched heterostructure, AJS, MJG, and JRP wrote the manuscript with input from all the authors.

\section{Competing Interests}

The authors declare no competing financial interests.

\section{Data Availability}

The data supporting the findings of this study are available within the paper and its Supplementary Information \cite{SOM}. The data are also available from the authors upon reasonable request.

\begin{acknowledgments}
Funded by Army Research Office grant No.\ W911NF-15-1-0149, DARPA grant No.\ D18AC0025, and the Gordon and Betty Moore Foundation's EPiQS Initiative through Grant GBMF4535. Devices were fabricated in the Princeton University Quantum Device Nanofabrication Laboratory. The authors acknowledge the use of Princeton’s Imaging and Analysis Center, which is partially supported by the Princeton Center for Complex Materials, a National Science Foundation MRSEC program (DMR-1420541)
\end{acknowledgments}

%

\end{document}